# A New Binary Star System of EW Type in Draco: GSC 03905-01870

BARQUIN, SALVADOR [1]

1) 39700 Castro-Urdiales, Cantabria, Spain, salvadorbarquin@outlook.com

**Abstract:** Discovery of a new binary star system (GSC 03905-01870 = USNO-B1.0 1431-0327922 = UCAC4 716-059522) in the Draco constellation is presented. It was discovered during a search for previously unreported eclipsing binary stars through the ASAS-SN database. The shape of the light curve and its characteristics (period of 0.428988±0.000001 d, amplitude of 0.34±0.02 $V$ Mag, primary minimum epoch HJD 2457994.2756±0.0002) indicates that the new variable star is an eclipsing binary of W Ursae Majoris type. I registered this variable star in The International Variable Star Index (VSX), its AAVSO UID is 000-BMP-891.

First, I decided to choose an area with a radius of 200 arcseconds around the star cataloged as TYC 3905-2030-1 (RA(2000) = $18^h35^m43.111^s$, DEC(2000) = +53°08'56.79") in the Tycho-2 Catalogue (Høg et al., 2000), located in the Draco constellation. Then, the UCAC4 Catalogue (Zacharias et al., 2012) was consulted through the VizieR database (Ochsenbein et al., 2000) in order to obtain the list of stars located in the area. After that, only those stars brighter than 15 mag in $V$ were selected (Tab. 1) to check their variability, with the ultimate goal of discovering a binary star system.

Table 1: Stars selected from the UCAC4 Catalogue whose variability was checked

| ID (UCAC4 Catalog) | Other identifications | RA [J2000] | DEC [J2000] | $V$ Mag |
|---|---|---|---|---|
| UCAC4 716-059507 | GSC 03905-02030　TYC 3905-2030-1 | $18^h35^m43.107^s$ | +53°08'56.70" | 11.730 |
| UCAC4 716-059493 | GSC 03905-01146　TYC 3905-1146-1 | $18^h35^m24.144^s$ | +53°08'32.89" | 11.948 |
| UCAC4 716-059500 | GSC 03905-01114　2MASS J18353330+5307054 | $18^h35^m33.300^s$ | +53°07'05.48" | 13.540 |
| UCAC4 716-059502 | GSC 03905-01190　2MASS J18353608+5309015 | $18^h35^m36.082^s$ | +53°09'01.62" | 13.839 |
| UCAC4 716-059509 | GSC 03905-00596　2MASS J18354504+5307568 | $18^h35^m45.042^s$ | +53°07'56.96" | 14.404 |
| UCAC4 716-059522 | GSC 03905-01870　2MASS J18355880+5310289 | $18^h35^m58.793^s$ | +53°10'28.92" | 14.637 |

In order to check the variability of the stars, their light curves were retrieved from the ASAS-SN database (Kochanek et al., 2017; Shappee et al., 2014) querying through its website. After that, the analyses of periods were performed using the ANOVA algorithm (Schwarzenberg-Czerny, 1996) implemented in the software PERANSO 2.51 (Paunzen & Vanmunster, 2016).

As a result of this, only one star cataloged as UCAC4 716-059522 (RA(2000) = $18^h35^m58.793^s$, DEC(2000) = +53°10'28.92") was clearly a variable star regarding its light curve. The retrieved data for this star covered 896 data points from HJD 2456037.1285 to 2458189.0739 with a mean photometric error of 0.043 *V* Mag.

The position of the variable star was queried in the The International Variable Star Index (Watson et al., 2006), where it was still unreported. The star also did not appear in Simbad (Wenger et al., 2000) or in the Tycho-2 Catalogue.

Having already previously determined the period P = 0.428988±0.000001 days, the polynomial fit algorithm implemented in PERANSO 2.51 was then used to determine the primary minimum epoch HJD 2457994.2756±0.0002 (through a routine of order 15).

Following that, I used the software VSTAR 2.19.0 (Benn, 2012) to create the phase plot (Fig. 1) that shows how the *V* Magnitude varies during the orbital cycle. This phase plot, made with data from ASAS-SN, shows clearly that the type of the star is a W Ursae Majoris variable. Additionally, I estimated the magnitudes of maximum (Max), primary minimum (Min I) and secondary minimum (Min II) using the polynomial fitting algorithm tool implemented in VSTAR 2.19.0 giving as a result the magnitude values listed in Tab. 2.

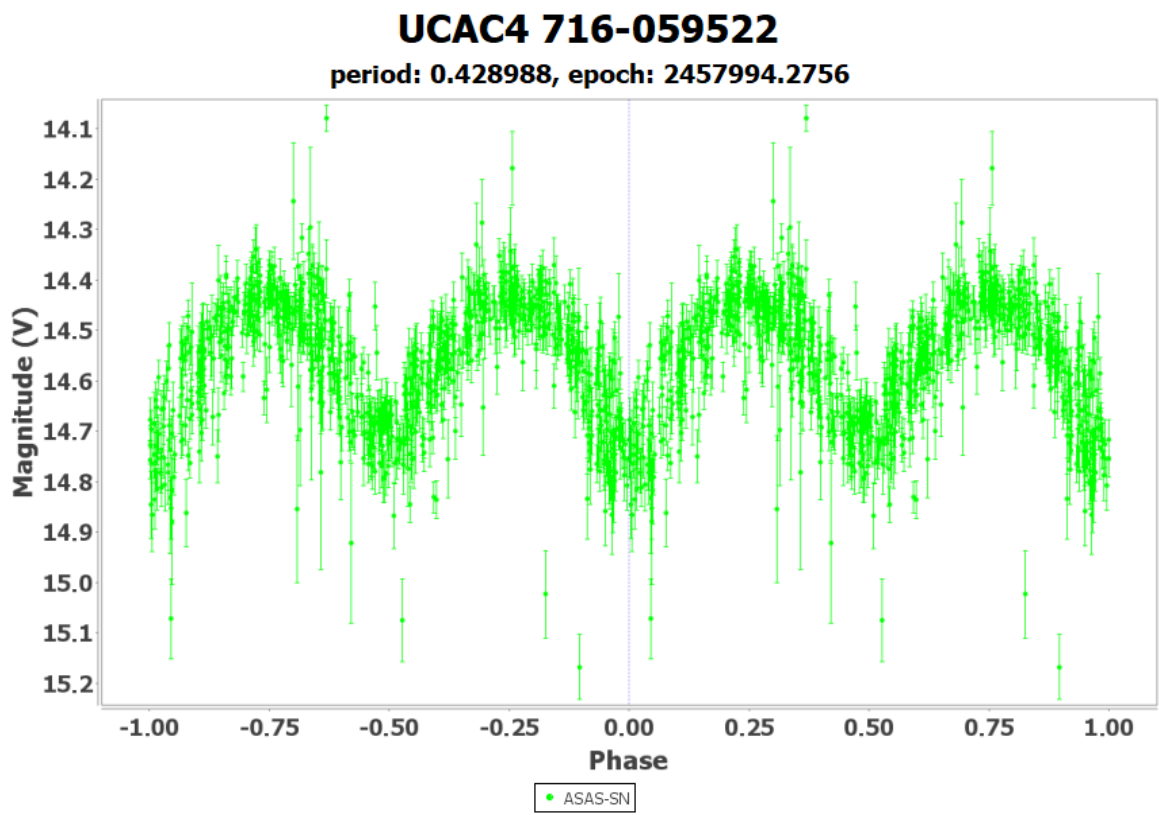

Figure 1: Phase plot made with data from ASAS-SN Database for period P = 0.428988 days and epoch HJD 2457994.2756

This new eclipsing binary star was reported to The International Variable Star Index, providing all the necessary data including cross-identifications from existing catalogs. Tab. 2 shows all the information and basic parameters of the star. This variable star was assigned the AAVSO UID 000-BMP-891 after its approval.

Table 2: Essential data of the variable star GSC 03905-01870

| ID (Guide Star Catalog) | GSC 03905-01870 |
|---|---|
| Other identifications | UCAC4 716-059522<br>USNO-A2.0 1425-09213580<br>USNO-B1.0 1431-0327922<br>2MASS J18355880+5310289 |
| Right Ascension [J2000] | $18^h35^m58.793^s$ |
| Declination [J2000] | +53°10'28.92" |
| Position source | UCAC4 Catalogue |
| Magnitude range (*V*) | 14.43 - 14.77 |
| Period (d) | 0.428988 ±0.000001 |
| Epoch (HJD) | 2457994.2756 ±0.0002 |
| Variability type | EW |
| Max magnitude (*V*) | 14.43 ±0.01 |
| Min I magnitude (*V*) | 14.77 ±0.01 |
| Min II magnitude (*V*) | 14.72 ±0.01 |

Acknowledgements: This research has made use of the SIMBAD and VizieR databases operated by the Centre de Données Astronomiques (Strasbourg). I would like to also thank Sebastián A. Otero (VSX-AAVSO) and OEJV editorial board for their helpful assistance.